\documentclass[12pt,osajnl2,preprint,showpacs,supersciptaddress]{revtex4}

\usepackage{graphicx}
\usepackage{subfigure}
\usepackage{dcolumn}
\usepackage{float}
\usepackage{amsmath}

\newcommand{\comment}[1]{}

\begin{document}

\title{\boldmath Flashes of light below the dripping faucet: an optical signal from capillary oscillations of water drops.\unboldmath
}

\author
{T. Timusk,$^{1,2}$ }

\email{timusk@mcmaster.ca.} 

\affiliation{$^{1}$Department of Physics and Astronomy, McMaster University, Hamilton, ON
L8S 4M1, Canada\\ $^{2}$The Canadian
Institute of Advanced Research, Toronto, Ontario M5G 1Z8, Canada}

\date{\today}

\begin{abstract}
Falling water drops from a dripping faucet, illuminated from above,  exhibit a row of bright strips of light, a few centimeters apart  at a fixed distance below the faucet.  Flash photographs of the drops show that they are oblate in shape when the flashes occur and the bright flashes of light originate from the edge of the drop that is on the opposite of the overhead light source. Here we show that that the spots result from the same internal reflection that gives rise to the rainbow in a cloud of spherical drops .  The periodic flashes reflect the capillary oscillations of the liquid drop between alternating prolate and oblate shapes and the dramatic enhancement in the oblate phase result from a combination of several optical effects. Ray tracing analysis  shows that the flashes occur when the rainbow angle,  which is 42 $\deg$ in spherical drops, but sweeps over a wide range between 35 $\deg$ and 65 $\deg$ for typical ellipsoidal drops and the intensity of the caustic is strongly enhanced in the oblate phase.  This phenomenon can be seen in all brightly lit water sprays with millimeter size drops and is responsible for their white color.  
\end{abstract}

\pacs{010.0010,290.1350}

\maketitle

\section{Introduction}

The study of the optical properties of water drops has a venerable history.  Centered around the understanding of atmospheric phenomena such as the rainbow, the problem has attracted the interest of many pioneers in optics starting with Descartes who first showed that the rainbow resulted from the reflection of sunlight on the inside surface of a spherical water drop.\cite{descartes}  This reflected beam is particularly strong when the angle between the incident rays and the observer is 42  $\deg$, the so-called rainbow angle, and the rays form a sharp bundle or caustic giving rise to a bright image.  In contrast with this internal reflection, narrowly concentrated in angle, light reflected from the outside of the drop, with the drop acting like a spherical mirror, is spread uniformly over all angles which results in a weak reflection.  If the drop diameter approaches the wavelength of light, diffraction effects become important and new phenomena come into play under the general rubric of Mie theory.\cite{bohren,marsden80}  Less studied are the optical properties of large drops where the shape of the drop is not confined by surface tension to be a sphere.  Such drops have irregular shapes governed by a balance between surface tension and air resistance for falling drops or gravity for drops that are attached to solid objects by surface tension. 

Water drops, such as the ones generated by a dripping faucet,  are 4 to 5 mm in diameter, and are to a first approximation spherical, but they can be deformed by capillary oscillations if they are subjected to an external disturbance. These oscillations were investigated theoretically by Rayleigh~\cite{rayleigh} who  showed that the n'th normal mode had a frequency given by $\omega$ (rad/sec) in

\begin{equation}
\omega_n^2={{n[n-1(n+2)]\sigma} \over {\rho a^3}}
\end{equation}
where $\sigma$ is the surface tension, 71.75 mN/m for water at 20 $\deg$ C, $\rho$ the density of water, 0.9982 g/ml at room temperature, and $a$ the radius of the drop. 
In the lowest mode, the quadrupole mode, $n=2$, the drop oscillates between prolate and oblate elliptical shapes with a frequency: 

\begin{equation}
\omega_2^2={{8\sigma} \over {\rho a^3}}.
\end{equation}

Capillary oscillations of water drops were first studied experimentally by Lenard~\cite{lenard} and their relevance to the optics of the rainbow have been summarized by Voltz.\cite{volz}   

The radius of a water drop from a dripping source is determined by a balance between gravity and surface tension just before it is released (Tate's law) resulting in drops that are weakly dependent on the size of nozzle,\cite{degennes} independent of the  flow rate.  We measured a diameter of 4.7 $\pm$ 0.4 mm in our experiments.  With this size, the calculated time for a full cycle of a quadrupole oscillation for  water at room temperature is 30 $\pm$ 4 ms.  This is in agreement  with the early measurements of Lenard. If the flow rate is increased to the point where the water forms a stream, a different mechanism of drop formation takes place~\cite{degennes} and the distance from the nozzle to the drop formation point varies with the flow rate.  Thus to study quadrupole oscillations, a slowly dripping source is preferred since not only is drop size constant, but the initial phase is well defined. The released drop is elongated and starts in the prolate phase.   

The kinematics of a millimeter size falling drop are fairly straightforward.  At small flow rates the drops are released with a small initial velocity of the order of 30 to 40 cm/s which depends on the flow rate and is related to momentum conservation between the flowing liquid in the pipe and the drop.  When formed, the drop it is in free fall until it reaches terminal velocity, about 1000 cm/s for millimeter size drops in air~\cite{bohren} at which point the drop is several meters below the release point.  The shape oscillations that are excited at the release point are underdamped, according to Chandrasekhar~\cite{chandrasekhar58} in the size range of a few millimeters, with a damping constant in sec$^{-1}$:

\begin{equation}
1/\tau=\nu(n-1)(2n+1)/a^2
\end{equation}
here $\nu$ is the kinematic viscosity, $\nu = 0.01004$ for water at 20 $\deg$ C.  For drops with $a=2.1$ mm we find for the lowest $n=2$ mode, $\tau = 1.0$ second, corresponding to a free fall distance of 500 cm.      

\section{Experimental results}

Figure 1 shows the geometry used to photograph the falling drops. The vertical axis is the direction of the velocity of the falling drop.  The drop is symmetric about this axis but distorts along this axis taking on alternate prolate and oblate shapes.  The light source is located at L at an angle $\theta$ above the horizontal plane and the camera at C at angle $\phi$ above the horizontal plane.  The angle $\alpha$ is the angle between the observer a C and the light source at L. Panel b) shows the path of  the rays inside the drop that form the strongest reflected image.  Rays enter the drop at a) and are reflected at the back surface at b) and emerge at c).  For a drop flattened along the vertical line (dashed line) the rays entering near the edge as shown can undergo total internal reflection at b) giving rise to an intense backscattered beam. 

Figure 2 shows a  sequence of stroboscopic images of  two falling water drops. The flow rate is such that a drop has moved out of the field of view at the bottom of the picture before the next drop is released from the faucet. This is shown clearly in panel b). The drops are illuminated with an off-camera flash (Nikon 8600), operating in a repetitive mode yielding flashes at intervals separated by 10.0 ms.  The orientation of the light and the camera are such that the flash angle $\theta$ is about 30 degrees the camera axis angle $\phi$  is 8 degrees. The camera to flash angle $\alpha$ is about 30 degress.  For best results the flash-to-camera angle should be from 30 to 50  degrees.  From a study of hundreds of  photographs such as the one in Fig. 1 one finds that initially, when the drop is released, it has an oblong, prolate shape as expected, and that it oscillates with a period of 30 ms between this shape and the pancake-like oblate form becoming spherical between these extremes.  The peak to peak amplitude of these quadrupole oscillations is small but can be measured on the photograph to be about 10 \% of the drop diameter. 

In panel a) the strobe light starts before the drop is released. According to previous studies~\cite{peregrine90,degennes} when the liquid bridge between the drop and the faucet is broken, the drop has a prolate shape and the stored elastic energy sets up quadrupole oscillations along the vertical axis.  Panel b) shows a sequence that starts once full cycle after the break of the bridge where the drop has returned to the prolate shape.  Half a cycle before this, as seen in panel a), the shape is oblate and a bright spot of light, characteristic signature of the oblate phase can be seen about 10 mm below the faucet. The second bright spot occurs 23 mm below the faucet. Panel (b) also shows faint images of small secondary droplets formed from the narrow neck of the liquid bridge.

We focus our attention on the very bright images seen at the lower left boundary of the drops in the oblate phase of the oscillations in Fig. 1.   In both pictures several bright images can be seen, separated by two darker drops in each case. Since the period of our strobe flash is 10 ms these bright images of the drops are approximately 30 ms apart, equal to the period of the lowest quadrupole mode of the capillary oscillation.  That these oscillations start with the release of the drop is clear from the spacial position of of the bright maxima, always at the same distance below the faucet.  The weak reflections of the flash unit towards the top-right in all the images are the reflection on the outer surface of the drop which is acting as a convex mirror.  This image is present in all the pictures of the drop as it falls including the completely spherical shapes.

The bright flashes of light can be also photographed by a much simpler arrangement  with a camera on a tripod, against dark background with a one second time exposure.  Figure 3 shows one such time exposure on the left along with a stroboscopic picture of the drops on the right. Here the incident beam is at $\theta=40$ deg and the camera at $\phi=35$. The scattering angle $\alpha=50$ deg.    Three strips of light can be seen centered at 8, 20.5 and 39 below the faucet.  It is also clear that the bright phase of the oscillation is quite long, lasting approximately for half of the full cycle and that the bright streak occurs  in the oblate phase.  If one moves the camera further from the source one finds that the bright phase gets shorter and shorter.  When the angle $\alpha$ exceeds a critical value, roughly 60 degrees the bright streaks disappear completely and the water stream is black.

To further test the hypothesis that these are capillary oscillations we repeated our experiment with methyl alcohol.  Methyl alcohol has a surface tension of only 22.6 mN/m.  The main effect of this is a smaller drop; we measured a diameter 3.6 mm.  It also has a lower density $\rho=0.792\times10^3$ Kg/m$^3$ at 20 $\deg$ C.  These factors largely cancel the lower surface tension in Rayleigh's formula resulting in a calculated period of 31.6 ms,  only slightly larger than for water drops.  Indeed, our experiments show that although the alcohol drops are much smaller than water drops, the bright images of methyl alcohol drops occur in the same place within the accuracy of our photographs as those of water drops.

\section{Discussion}

What causes the bright flash of light in the oblate phase of the drop?  The position of the light on the drop and its angular location relative to the flash suggests  that  the light originates from a single internal reflection, the same path that the primary rainbow image takes shown in Fig. 1b.  This is the $p=2$ term in the Debye series used to discuss diffraction phenomena in small drops. \cite{li07} To investigate in detail the effect of the non-spherical shape of the drop on the intensity of the reflected light we wrote a ray tracing program that uses geometrical optics to follow the path of 200,000  parallel rays incident on an ellipsoidal drop falling on random positions on the surface of the drop and with random polarizations. At each surface we calculated the reflected and transmitted electric fields for using Frensel's equations for $s$ and $p$ polarizations.   We sorted the $p=2$ outgoing rays into bins according to scattering angle $\alpha$, {\it i.e.} the angle between the light source and the observer and the angle around the circle.  Since the size parameter $x=2\pi a/\lambda$ in our drops is 4600, we feel justified to use geometrical optics limit of scattering to model the propagation of light inside the drop and our method is only valid in the limit of drop size that is much larger than the wavelength.\cite{bohren,wang91}  

Figure 4. shows the calculated  pattern of scattered light  from the $p=2$ reflection as a function of vertical and horizontal angle. The incident direction is 45 degrees above the horizontal plane and the symmetry axis of the drop is vertical.  The coordinates have been rotated in such a way as to place the incident direction in the center of the diagram.  A conventional rainbow, in this diagram, would be a circle with a radius of $\alpha =$42 degrees centered on the incident direction. It is shown as a dashed circle.   The back scattering from ellipsoidal drops forms elliptical "rainbows".  The curves are for eccentricities ranging from $\epsilon= -0.06$ to $+0.06$  where the axes of the ellipsoid are $(1+\epsilon,1+\epsilon,1-2\epsilon)$ with $\epsilon > 0$ for the oblate shape and $\epsilon < 0$ for the prolate shape. The major axis of oblate drops is horizontal and prolate drops vertical.  At 45 degree angle of incidence shown in the figure  the centers of the ellipses are displaced from the incident direction. Other angles of incidence also produce elliptical patterns. For example for light entering close to the symmetry axis the images are circles with larger radii in the oblate phase.  For light entering normal to the symmetry axis, {it i.e.} the horizontal plane for falling drops, the images are ellipses with the major axis vertical but centers displaced from the horizon. 

The scattering efficiency $Q$ is defined as the ratio of the scattered flux to incident flux on a particle.~ \cite{bohren} In 
Figure 5  we plot the scattering efficiency per unit solid angle {\it i.e.}  the differential scattering efficiency $Q'$ of  light scattered from ellipsoidal drops as a function of scattering angle $\theta$ along a great circle intersecting the symmetry axis of the drop {\it i.e.} the north pole of the flattened earth-like ellipsoidal drop.  The region of angles covered is shown in Figure 4. The incident angle is 45 $\deg$ above the horizontal plane.  For a spherical drop, the once internally reflected rays form a sharp peak at $\theta=42 \deg$ and there are no rays with $\theta>42.4 \deg$. All the curves are cut off at a maximum angle $\theta_{max}$. Angles larger than  $\theta_{max}$ are in what for conventional rainbows are called the Alexander's dark band. As we see, $\theta_{max}$ increases dramatically with increasing flatness of the drop from 42.4 degrees for a spherical drop to over 60 degrees for the most oblate drop.  Also, we see an increase in overall intensity of the reflected light with increasing flatness. For the spherical water drop $Q =0.04$, {\it i.e.} 4 \% of the incident energy on the ellipsoid is goes to the rainbow image.  We contrast this with the oblate ellipsoid at 45 degree angle of incidence and $\epsilon=0.06$ where $Q=0.052$, substantially larger.  However the efficiency is much stronger close at the angle of the caustic and extends over a much broader angle for the oblate shaped drop.  

Ray tracing also shows that the brightness of the  image seen in oblate drops is enhanced by a geometric effect caused by the elliptic shape of the drop which leads to an angle of incidence greater than $ \sin^{-1}(1/n)=48.6$ degrees  for $n=1.33$  for the reflection on the back surface of the drop. This is the condition for total internal reflection (TIR).  Thus an oblate drop, illuminated from above, gives rise to a bright image from TIR light  at the edge of the drop.  A spherical drop does not give rise to TIR images. Prolate drops can also give rise to TIR and these can be seen for incident angles close to the horizontal symmetry plane of the drop. 

Figure 5 also shows why the observer sees strips of light from a single oscillating drop. Situated at a fixed angle to the source, say where the angle of deflection $\alpha=50$ deg, as the drop changes its shape no light is seen when the drop is in Alexander's dark band. But whenl the eccentricity becomes positive a strong scattered beam is seen for $\epsilon > 0.03$.   Fig. 6 also shows that the length of the bright phase depends on the scattering angle getting shorter for large scattering angles.  The length of the bright phase in Fig. 3 is consistent with 50 deg scattering angle used in the experiment.

It is important to note that large water drops, drops that are larger than the wavelength of light, do not scatter much light.  Viewed from any angle except right into the light source they will appear black.  This can be seen clearly in fig. 2. Geometrical optics of a sphere shows that the reflected image from the front surface has 6.6 \% of the incident intensity while 88.4 \% goes to the transmitted beam.  The internally reflected beam, the one  that forms the rainbow, has only 4.0 \% intensity. These three rays account for 99.7 \% of the incident intensity deflected by a spherical drop~\cite{bohren}.  
 
Why then do water sprays appear white?  A close inspection of any such spray suggests the answer.  Observed visually with an overhead light source it is clear that the spray consists of short  strips of bright light.  A flash photograph of a bath shower, shown in Fig. 6,  shows that while most of the drops are indeed black as predicted by geometrical optics of a sphere, there are some very bright drops that provide the light.  The photograph also shows that while  most drops are spherical in shape many are quite irregular.  Nevertheless, the strips of bright light are more or less the same length and can be seen as far as 2 m  from the shower head.  The strips are not seen in diffuse light confirming the idea that the scattering is caused by geometrical optics of the drops and not multiple isotropic scattering as seen in clouds.  A typical spray is optically thin.  It is clear that the same phenomenon of capillary oscillations of droplet shape giving rise to strips of light below the dripping faucet is favorable towards strong scattering of light in all sprays .  It is surprising that this common phenomenon is rarely discussed in the literature although we found a photograph of rainfall in ref. (4) where the author does interpret the bright traces due to capillary oscillations and estimates the drop size using Rayleigh's formula.  But as Fig. 6 shows the location of the bright image varies from drop to drop which makes any quantitative study of the optics of sprays difficult.  In a spray the oscillations are driven by turbulence and are a complex combination of rocking and zig-zag motions combined with the capillary oscillations\cite{clift}.   In contrast, the dripping tap provides a system with well defined starting conditions for the oscillations. 

Finally returning to the rainbow one has to note that it is generally assumed that falling raindrops that are large will be flattened due to air resistance and undergo capillary oscillations.\cite{volz}. However as the present study shows, such drops cannot give rise to a conventional rainbow. The p=2 caustic varies in angle from 30 to 60 degrees in the course of one cycle completely washing out any wavelength dependent dispersion that requires stability of the rainbow angle to better than one degree.  We conclude that only very small, non-oscillating drops can give rise to colorful rainbows.    

In summary, experiments show that large water drops scatter light strongly in the oblate phase of capillary oscillations.  We suggest that this effect is also responsible for the strips of light seen in most sprays and ultimately for their white color when illuminated by a fairly concentrated overhead source.  We also suggest that any realistic computer simulation of water sprays be in the form of bright strips of light in the direction of the flow of the spray.  Another application of the phenomenon is to Fourier transform the light scattered by the spray and use the resulting peak at the Rayleigh frequency to determine average drop size optically.

\section*{Acknowledgements}

This work has been supported by the Canadian Natural Science and Engineering Research
Council and the Canadian Institute of Advanced Research. We would like to acknowledge valuable suggestions by Kari Dalnoki-Veress and  technical help by Andy Duncan and Greg Egan.


\begin{figure}[t]
\includegraphics[width=15cm]{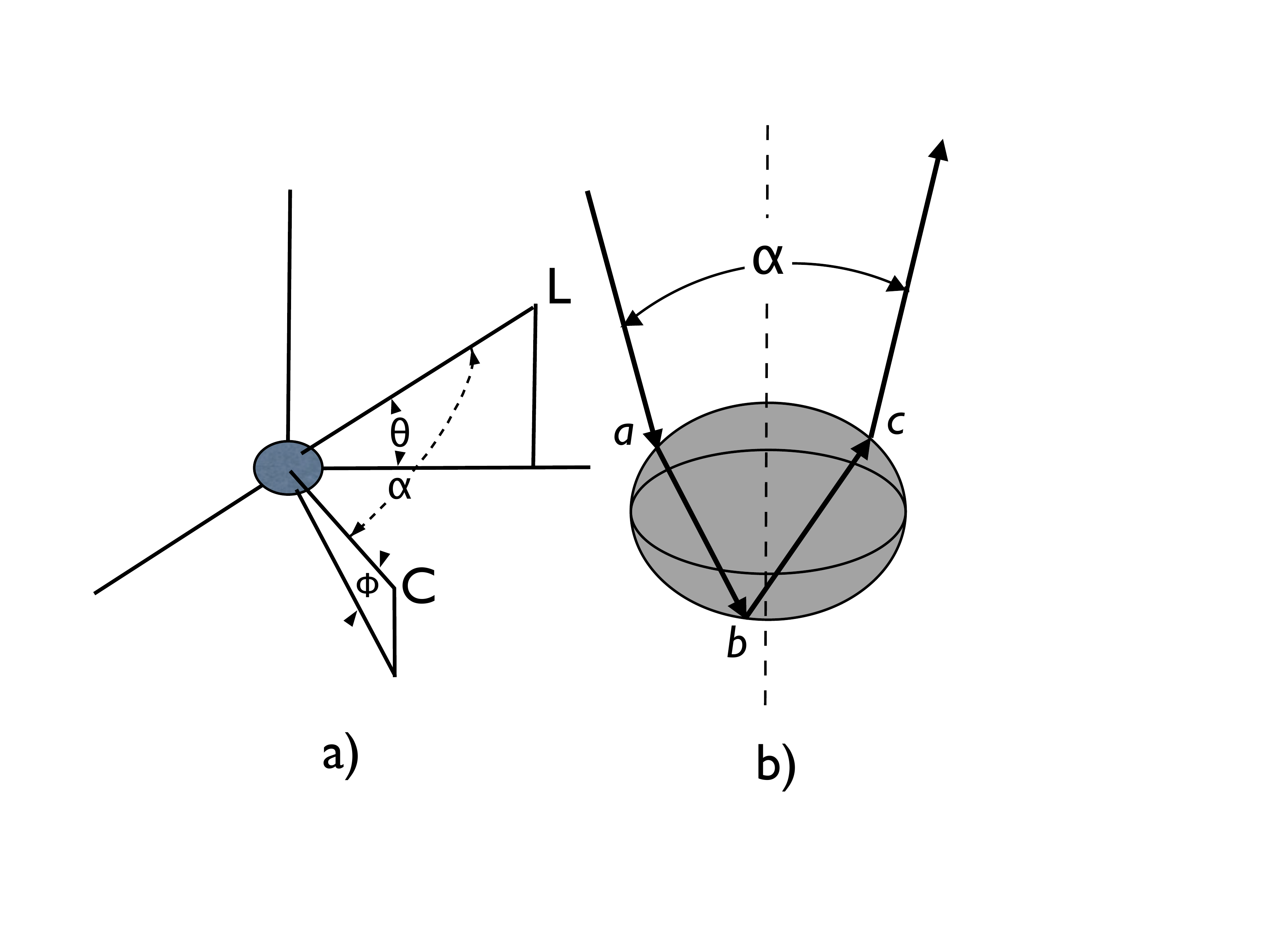}

\caption{(Color online) Geometry used to photograph falling drops.  Panel a) shows the ellipsoidal drop at the origin with its symmetry axis vertical.  The light source is located at L at an angle $\theta$ above the horizontal.  The camera at C is at an angle $\phi$ above the horizontal plane.  The scattering angle $\alpha$ is the angle between the light source and the observer at C. Panel b) shows the rays propagating inside the drop. }
\end{figure} 

\newpage

\newpage
\begin{figure}[t]

\includegraphics[width=20cm]{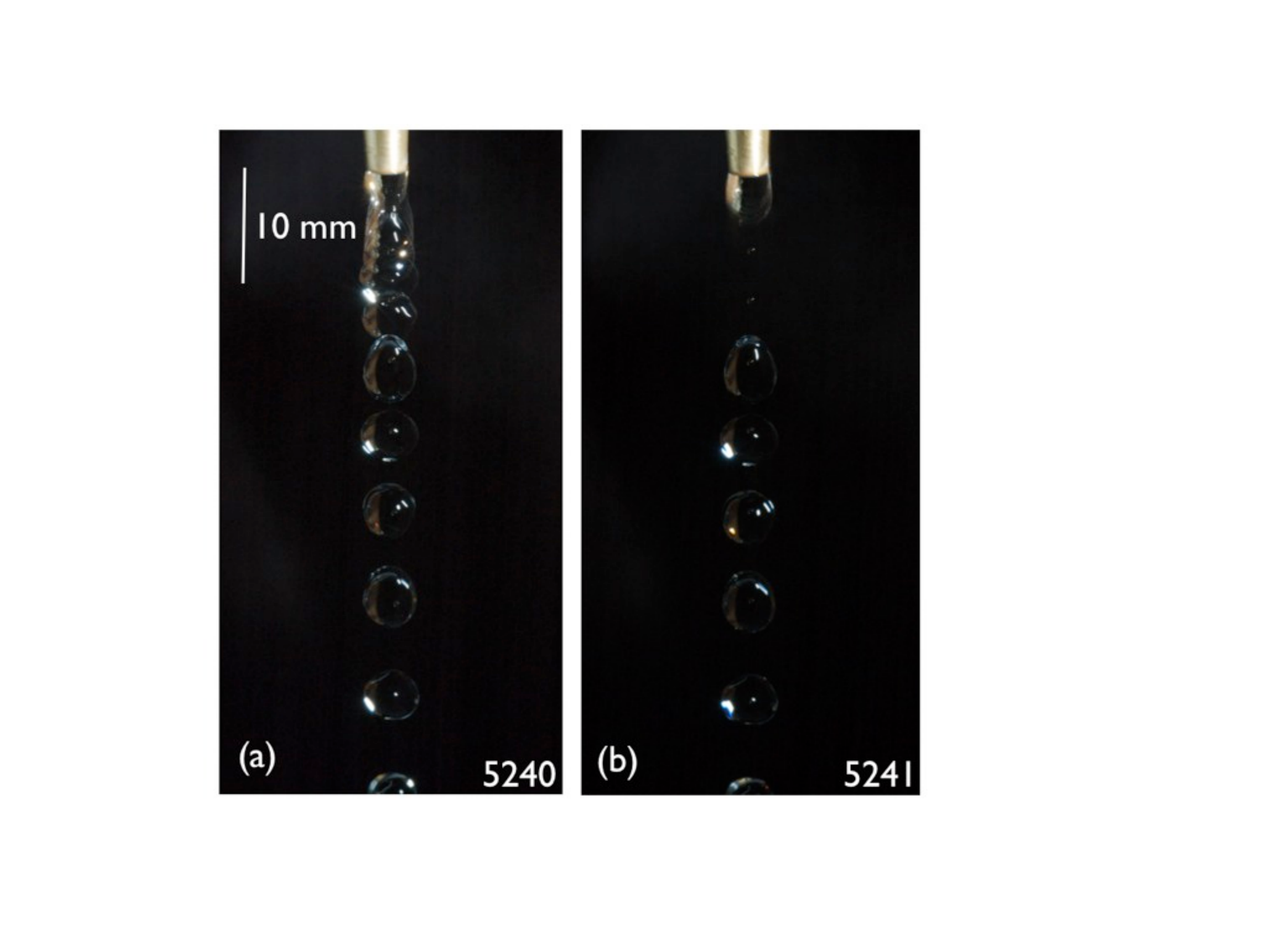}

\caption{(Color online) Stroboscopic photo of two water drops from a dripping faucet.  Images are 10 ms apart.  (a) shows a sequence of photos that start at the time of release of the drop and (b) a sequence that starts somewhat later.   We show that the bright spots seen on the lower left of certain drops are the result of total internal reflection of the light source (a camera flash in the stroboscopic mode) incident from the upper right.  The water drops that show the bright spots are oblate in shape.  The intermediate, dark drops are either spherical or prolate.   The elapsed time between the oblate phases is 30 ms, consistent with the calculated time for capillary shape oscillations of the drop.}
\end{figure}

\newpage
\begin{figure}[t]

\includegraphics[width=25cm]{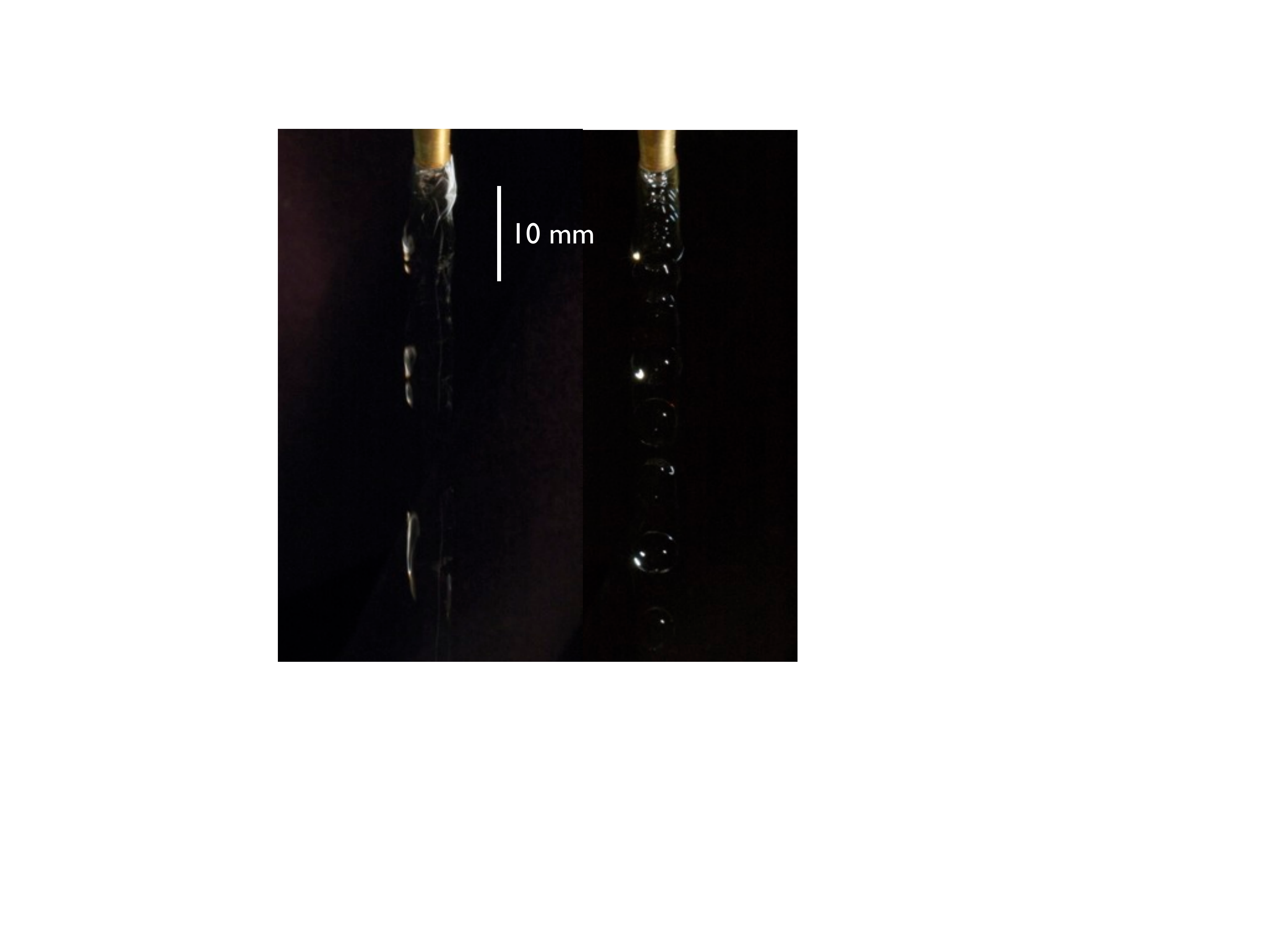}

\caption{(Color online) Time exposure of a water drop on the left and a stroboscopic image of a drop under the same conditions of flow rate on the right.}
\end{figure} 

\newpage
\begin{figure}[t]

\includegraphics[width=20cm]{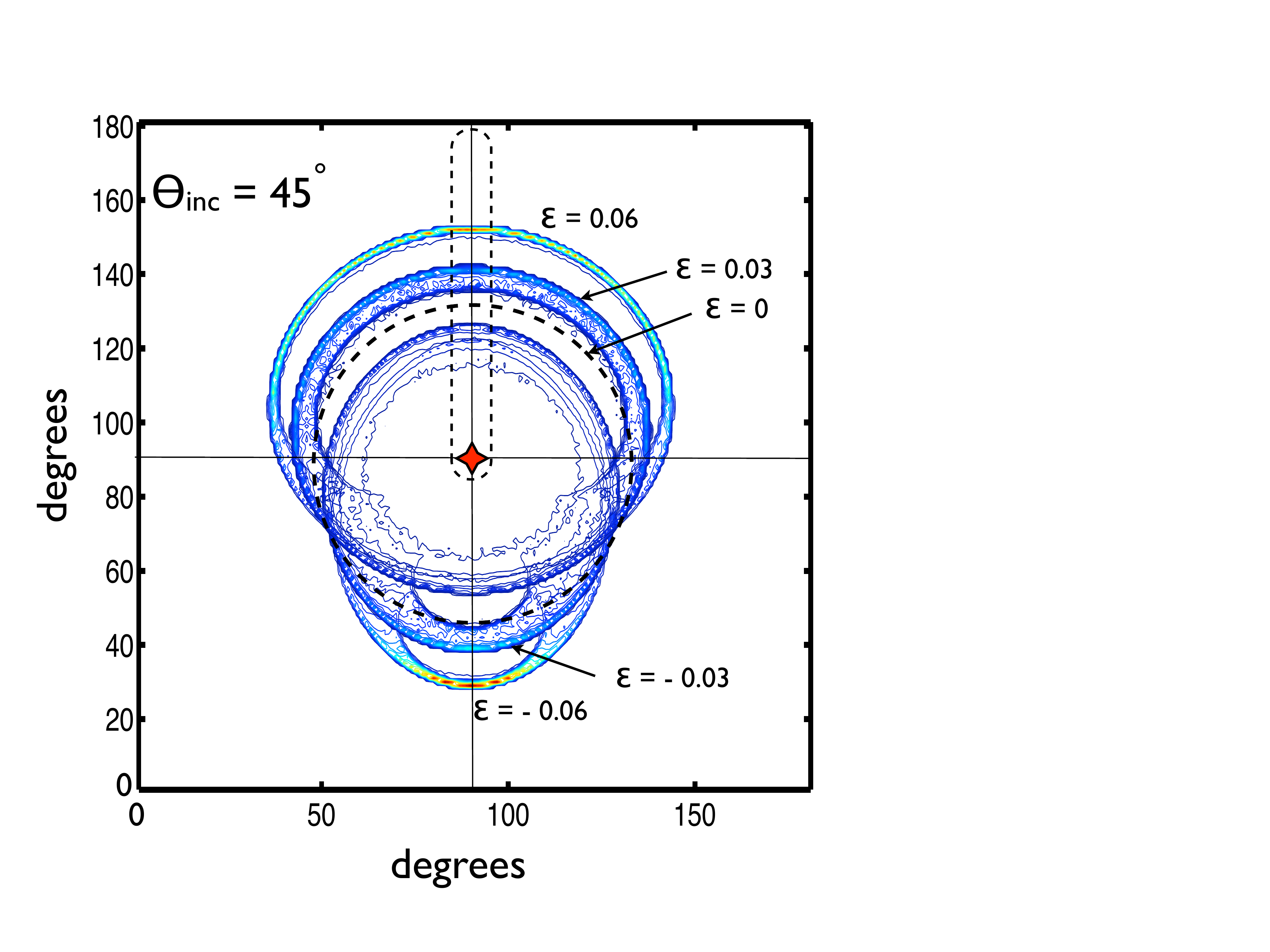}

\caption{(Color online) "Rainbow" images of backscattering from ellipsoidal water drops.  Light is incident at 45 degrees from the vertical symmetry direction of the drop.  Various eccentricities are shown.  The dashed circle is where the rainbow from spherical drops is located, 42 degrees from the incident direction marked by a star at the center of the figure.  Ellipsoidal drops give rise the elliptically  shaped rainbows as shown.  However a rapidly oscillating drop would show elliptical patterns that change in size and shape as seen in the figure.  The oblate ellipsoid gives rise to a particularly well defined ellipsoid. As the shape approaches this  form the scattered efficiency increases dramatically in strength and angular coverage, up to 65 degrees from the incident direction.  The dashed vertical box shows the region of angles that were included in Fig. 5}
\end{figure}

\newpage
\begin{figure}[t]

\includegraphics[width=20cm]{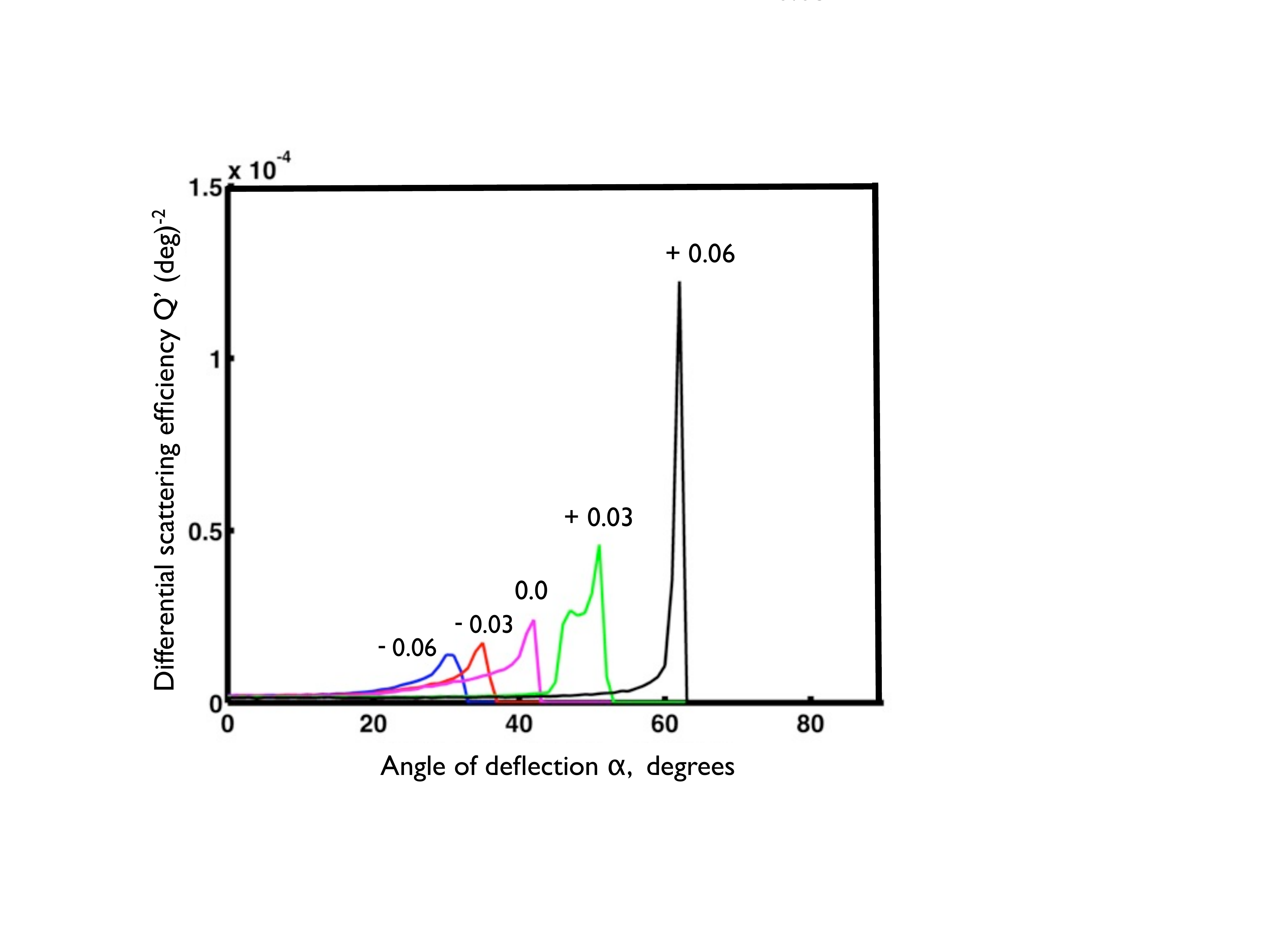}

\caption{(Color online) Differential scattering efficiency of light $Q'$ scattered from ellipsoidal drops, plotted as a function of the angle from the incident direction (45 degrees above the horizontal plane) in a vertical direction.  From left to right curves range in eccentricity $\epsilon$ from $\epsilon =-0.06$ (prolate ellipsoid) to sphere $\epsilon =-0.00$ to oblate ellipsoid $\epsilon = +0.06$. As the shape approaches the oblate ellipsoid form the scattered efficiency increases dramatically in strength and angular size. }
\end{figure}

\newpage
\begin{figure}[t]

\includegraphics[width=5 in]{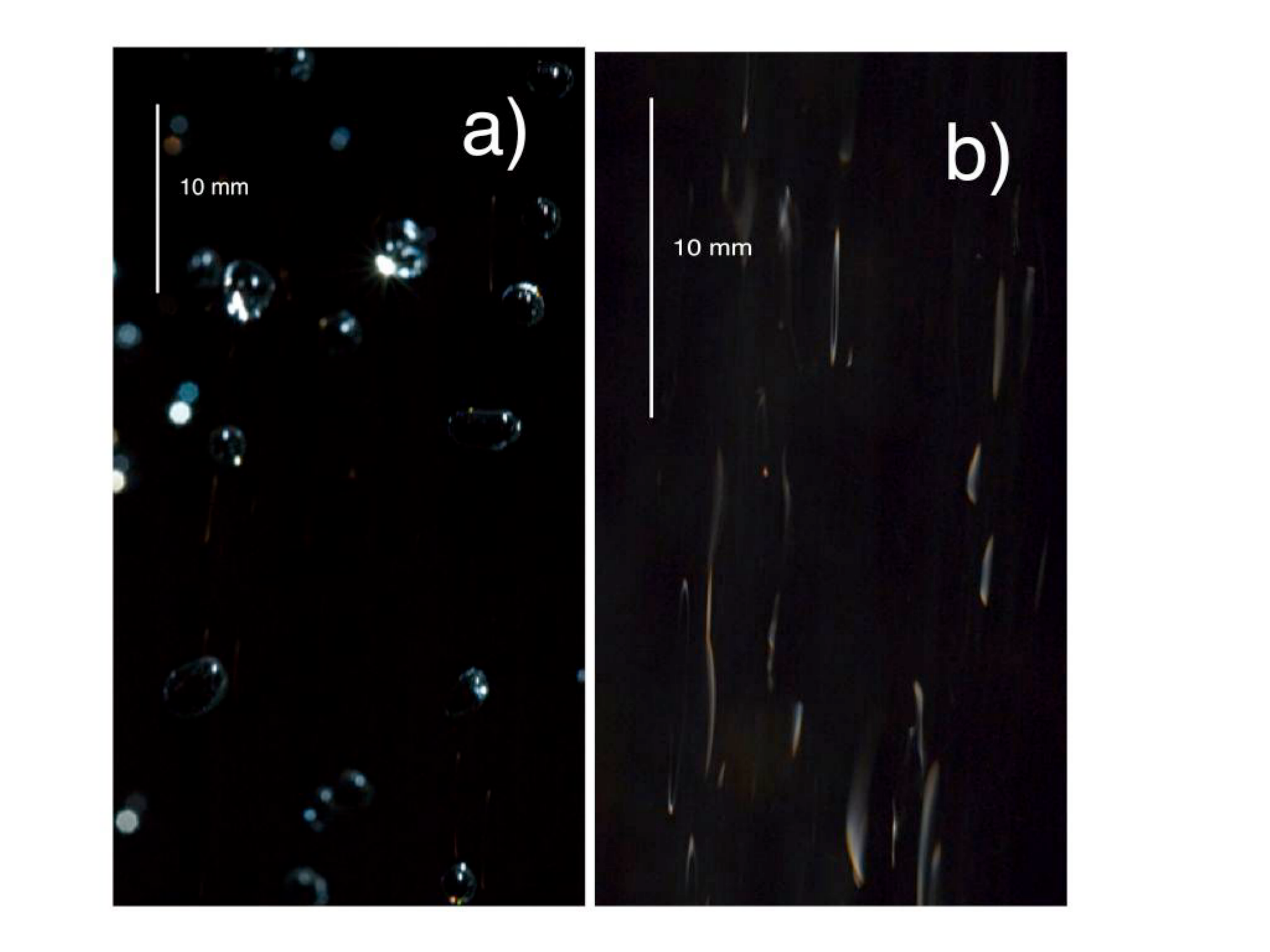}

\caption{(Color online) a) Flash photograph of a water spray from a collection of 1 mm diameter nozzles delivering drops with an initial velocity of 300 cms$^{-1}$. The flash is held 45 degrees from the camera's optic axis. Note the irregular shape  of the larger drops.  Some of the drops show bright flashes of light near the edge of the drop.   b) A photograph of the same spray as in a) but using an incandescent source and a time exposure.  Strips of light can be seen.  Their oval shape suggests they arise from the internal reflection of light where two virtual images of the source merge at the ``rainbow" angle of maximum deviation.  These strips are a few centimeters apart consistent with the notion that they related to capillary oscillations of the mm size drops seen in a).   
}

\end{figure}


\begin{thebibliography}{99}
\bibitem{descartes} Ren\'{e} Descartes, {\it Discourse of the Method} (1637).
\bibitem{bohren} C.E. Bohren and D.R Huffman {\it Absorption and scattering or light by small particles,} (Wiley-Interscience, 1983). 
\bibitem{marsden80} P.L. Marston, "Rainbow phenomena and the detection of nonsphericity in drops," Applied Optics, {\bf 19,} 680-689 (1980).
\bibitem{rayleigh} Rayleigh, Phil. Mag. (5) xxxiv. {\bf 177} (1892), H. Lamb {\it Hydrodynamics}, 6th. ed. (Cambridge University Press (1932) p. 473.  
\bibitem{lenard} Ph. Lenard, ``\"Uber die Schwingungen fallender Tropfen,"  Ann. Phys. Chem., {\bf 30}, 209 (1877).
\bibitem{volz} F.E. Volz, "Some aspects of the optics of the rainbow and the physics of rain," {\it Physics of Precipitation} ed. by H. Weickmann, American Geophysical Union monograph No. 5,  280-286 (1960).  
\bibitem{degennes} P-G. de Gennes, F.  Brochard-Wyart, D. QuŽrŽ, A. Reisinger {\it Capillarity and Wetting Phenomena: Drops, Bubbles, Pearls, Waves}, Springer, New York, 2004.
\bibitem{chandrasekhar58}  S. Chandrasekhar, ``The oscillations of a viscous liquid globe," Proc. London Math. Soc. {\bf 9}, 141- 149 (1959). 
\bibitem{peregrine90} D.H. Peregrine, G. Shoker, A. Symon, ``The bifurcation of liquid bridges," J. Fluid Mech. {\bf 212.} 25-39 (1990).
\bibitem{li07} Renxian LI, XiangÕe Han, Lijuan Shi, Kuan Fang Ren, and Huifen Jiang, ``Debye series for Gaussian beam scattering by a multilayered sphere", App. Optics {\bf 46,} 4808 (2007).  
\bibitem{wang91} Ru T. Wang and H. C. van de Hulst,  App. Optics {\bf 30} 106 (1991), ``Rainbows: Mie computations and the Airy approximation" 
\bibitem{clift} R. Clift, J.R. Grace and M.E. Weber, {\it Bubbles drops and Particles}, Academic Press, New York 1978, p. 185.

\end{thebibliography}
\end{document}